\documentclass[epjST]{svjour}
\usepackage{latexsym}
\usepackage{graphics}
\usepackage{graphicx,amssymb,epsfig,epsf,amsmath}
\usepackage{physics}

\def\be{\begin{equation}}
\def\lan{\left\langle}
\def\ran{\right\rangle}
\def\ee{\end{equation}}
\def\barr{\begin{array}}
	\def\earr{\end{array}}

\def\l{\left}
\def\r{\right}
\def\dis{\displaystyle}
\def\ed{\end{document}}

\def\cac{{\cal C}}

\def\can{{\cal N}}

\begin{document}
\title{Eigenstate structure in many-body bosonic systems: Analysis using random matrices and $q$-Hermite polynomials}
\subtitle{Eigenstate structure in many-body bosonic systems}
\author{Priyanka Rao$^1$ \and Manan Vyas$^2$ \and N. D. Chavda$^1$}
\institute{Department of Applied Physics, Faculty of Technology and Engineering, The Maharaja Sayajirao University of Baroda, Vadodara-390001, India \and Instituto de Ciencias F{\'i}sicas, Universidad Nacional Aut{\'o}noma de M\'{e}xico, 62210 Cuernavaca, M\'{e}xico 
}                     
%
%
%
\date{Received: date / Revised version: date}
%
\abstract{We analyze the structure of eigenstates in many-body bosonic systems by modeling the Hamiltonian of these complex systems using Bosonic Embedded Gaussian Orthogonal Ensembles (BEGOE) defined by a mean-field plus $k$-body random interactions. The quantities employed are the number of principal components (NPC), the localization length ($l_H$) and the entropy production $S(t)$. The numerical results are compared with the analytical formulas obtained using random matrices which are based on bivariate $q$-Hermite polynomials for local density of states $F_k(E|q)$ and the bivariate $q$-Hermite polynomial form for bivariate eigenvalue density $\rho_{biv:q}(E,E_k)$ that are valid in the strong interaction domain. We also compare transport efficiency in many-body bosonic systems using BEGOE in absence and presence of centrosymmetry. It is seen that the centrosymmetry enhances quantum efficiency.
} 
\maketitle

\section{Introduction}
\label{intro}

Random matrix theory (RMT) was introduced in physics by Wigner \cite{Wi-55} and the tripartite classification of classical random matrices (Gaussian Orthogonal Ensemble (GOE), Gaussian Unitary Ensemble (GUE) and Gaussian Symplectic Ensemble (GSE)) was given by Dyson \cite{Dy-62}. RMT is a multidisciplinary research area with numerous applications in various areas of science \cite{kota-book,SenRMP,Rigol,zel-lea,KC-18,Cotler-2017}. Constituents of isolated quantum systems interact via few-body interactions whereas the classical random matrix ensembles take into account many-body inetractions. Thus, random matrix models accounting for few-body interactions, called {\it Embedded ensembles} (EE), were introduced \cite{FW70,BF71}. 

$q$-Hermite polynomials have been employed to study spectral densities of the so-called SYK model \cite{GaVerb2016,GaVerb2017} and quantum spin glasses \cite{spin-g} alongwith studying the local density of states (LDOS) and survival probabilities in fermionic as well as bosonic EE \cite{Manan-Ko}. Formulas for parameter $q$ for fermionic and bosonic EE are derived in \cite{Manan-Ko} which explain the semi-circle to Gaussian transition in spectral densities and LDOS and the survival probability decay in many-body quantum systems as a function of rank of interactions. In the present paper, we analyze the structure of eigenstates using NPC, localization lengths ($l_H$) and entropy production $S(t)$. These three quantities measure delocalization in an isolated many-body quantum system. We model our system Hamiltonian by bosonic EE with mean-field interactions plus $k$-body random interactions, denoted as BEGOE($1+k$) (with $k \le m$, $m$ being the number of bosons distributed in $N$ single particle levels). We also analyze the effect of centrosymmetry on transport efficiencies. Now, we will give a preview.

For sake of completeness, BEGOE($1+k$) is briefly introduced and formula for parameter $q$ defining $q$-Hermite polynomials for BEGOE($k$) is given in Section \ref{sec:2} with an example for Gaussian to semi-circle transition in spectral densities. Results for NPC, $l_H$ and entropy production are described in Section \ref{sec:4}. Section \ref{sec:5} compares the distribution of the transport efficiencies in BEGOE(1 + $k$) without and with centrosymmetry.  Finally, Section \ref{sec:6} gives conclusions.


\section{Preliminaries}
\label{sec:2}

\subsection{BEGOE($1+k$)}

Consider $m$ spinless bosons distributed in $N$ degenerate single particle (sp) states interacting via $k$-body ($1 \leq k \leq m$) interactions. Distributing these $m$ bosons in all possible ways in $N$ sp levels generates many-particle basis of dimension $d_m={N+m-1 \choose {N}}$. The $k$-body random Hamiltonian $V(k)$ is defined as,
\be
V(k) = \dis\sum_{k_a,k_b} V_{k_a,k_b} B^\dagger(k_a) B(k_b)\;.
\label{eq.ent1}
\ee
Here, operators $B^\dagger(k_a)$ and $B(k_b)$ are $k$-boson creation and annihilation operators. They obey the boson commutation relations. $V_{k_a,k_b}$ are the symmetrized matrix elements of $V(k)$  in the $k$ particle space with the matrix dimension being $d_k={N+k-1 \choose k}$. They are chosen to be randomly distributed independent Gaussian variables with zero mean and unit variance, in other words, $k$-body Hamiltonian is chosen to be a GOE. BEGOE($k$) is generated by action of $V(k)$ on the many-particle basis states. Due to $k$-body nature of interactions, there will be zero matrix elements in the many-particle Hamiltonian matrix, unlike a GOE. By construction, we have a GOE for the case $k = m$. For further details about these ensembles, their extensions and applications, see \cite{kota-book,BRW,Manan-thesis,Ma-Th} and references therein.

In realistic systems, bosons also experience mean-field generated by presence of other bosons in the system and hence, it is more appropriate to model these systems by BEGOE($1+k$) defined by,
\be
H= h(1) + \lambda V(k)
\label{eq.ent2}
\ee
Here, the one-body operator $h(1)=\sum_{i=1}^N \epsilon_i n_i$ is described by fixed sp energies $\epsilon_i = i + 1/i$; $n_i$ is the number operator for the $i$th sp state. The parameter $\lambda$ represents the strength of the $k$-body ($2 \le k \le m$) interaction and it is measured in units of the average mean spacing of the sp energies defining $h(1)$.

A very significant property of BEGOE($k$) (and BEGOE(1+$k$)) is that in general they exhibit Gaussian to semi-circle transition in the eigenvalue density as $k$ increases from $1$ to $m$ \cite{MF}. This result is now well established from many numerical calculations and analytical proofs via lower order moments \cite{kota-book,KC-18,BRW,Asaga,Manan-1,SM}. Very recently, it is shown that generating function for $q$-Hermite polynomials describes the semi-circle to Gaussian transition in spectral densities and local density of states (LDOS) [also known as strength functions] alongwith survival probability decay using $k$-body EE (for the Orthogonal and Unitary variants) for fermionic and bosonic systems as a function of rank of interactions $k$ \cite{Manan-Ko}. In the next section, we give the formula for parameter $q$ for BEGOE($k$).

\subsection{BEGOE($k$): parameter $q$ defining $q$-Hermite polynomials}

L. J. Rogers first introduced the $q$-Hermite polynomials in mathematics \cite{ISV-87}. Consider $q$ numbers $[n]_q$ defined as $\l[n\r]_q = (1-q)^{-1}(1-q^n)$. Then, $[n]_{q \rightarrow 1}=n$, and $[n]_q!=\Pi^{n}_{j=1} [j]_q$ with $[0]_q!=1$. Now, $q$-Hermite polynomials $H_n(x|q)$  are defined by the recursion relation \cite{ISV-87}, 
\be
x\,H_n(x|q) = H_{n+1}(x|q) + \l[n\r]_q\,H_{n-1}(x|q)
 \ee
with $H_0(x|q)=1$ and $H_{-1}(x|q)=0$. Note that for $q=1$, the $q$-Hermite polynomials reduce to normal Hermite polynomials (related to Gaussian) and for $q=0$ they will reduce to Chebyshev polynomials (related to semi-circle). Importantly, $q$-Hermite polynomials are orthogonal within the limits $\pm 2/\sqrt{1-q}$, with the weight function $v(x|q)$ defined by \cite{Manan-Ko},
\be
\barr{rcl}
v(x|q) & = & \can_q\,\dis\sqrt{1-\dis\frac{x^2}{x_0^2}}\;
\dis\prod^{\infty}_{\kappa = 1} \l[
1-\dis\frac{4(x^2/x_0^2)}{2+q^\kappa + q^{-\kappa}}\r]\;;\;\;
x_0^2 = \dis\frac{4}{1-q}\;.
\earr
\label{eq.ent7}
\ee
Here, $x$ is standardized variable (centroid zero and variance unity), $-2/\sqrt{1-q} \leq x \leq 2/\sqrt{1-q}$ and $\can_q$ is the normalization constant. It is seen that in the limit $q \rightarrow 1$, $v(x|q)$ will take Gaussian form and in the $q=0$ limit semi-circle form.

The formula for $q$ for BEGUE($k$) [also valid for BEGOE($k$)] is given by \cite{Manan-Ko},
\be
\barr{l}
q \sim \dis\binom{N+m-1}{m}^{-1} \dis\sum_{\nu=0}^{\nu_{max}}\; \dis\frac{
X(N,m,k,\nu) \;d(g_\nu)}{\l[\Lambda^0(N,m,k)\r]^2} \;;\;\; \\ \\
X(N,m,k,\nu) = \Lambda_B^\nu(N,m,m-k)\;\Lambda^\nu(N,m,k)\;;\\ \\
\Lambda^\nu(N,m,r) =  \dis\binom{m-\nu}{r}\;\dis\binom{N+m+\nu-1}{r}\;,\\ \\
d(g_\nu)  = \dis\binom{N+\nu-1}{\nu}^2-\dis\binom{N+\nu-2}{\nu-1}^2\;.
\earr \label{eq.ent9}
\ee

\subsection{Spectral density}

\begin{figure}
		\begin{center}
			\includegraphics[width=0.9\linewidth,height=2.75in]{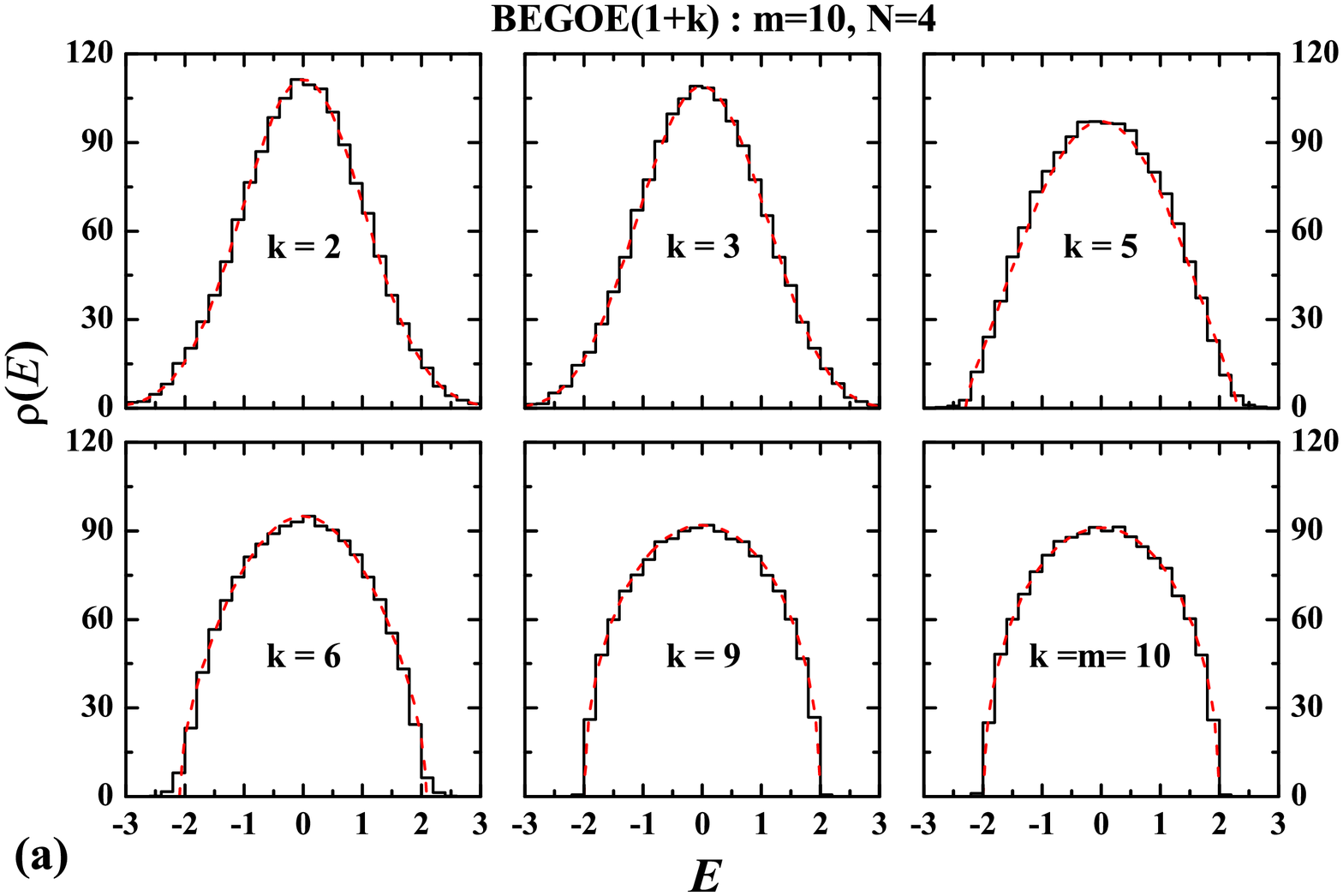}
		\end{center}
		\caption{Ensemble averaged spectral density $\rho(E)$ for a 100 member BEGOE($1+k$) ensemble with $m = 10$ bosons in $N = 4$ sp states for different $k$ values. We choose $\lambda = 0.5$. The dashed (red) curves are obtained using Eq.\eqref{eq.ent7} with the corresponding values of $q$ given by Eq.\eqref{eq.ent9}.}
\label{fig.rho}
	\end{figure}

We demonstrate the transition in spectral densities from Gaussian to semi-circle with increasing $k$ for a system of $m = 10$ bosons distributed in $N = 4$ sp levels in Figure \ref{fig.rho}. Numerical results (histograms) are for a 100 member BEGOE($1+k$) ensemble with $\lambda = 0.5$ and the dashed (red) curves are obtained using Eq.~\eqref{eq.ent7} with $q$ computed by Eq.~\eqref{eq.ent9}. First, the spectral density for each member of the ensemble is zero centred and scaled to unit width and then the histograms are constructed. The numerical results clearly display transition in the spectral density from Gaussian to semi-circle form as $k$ changes from 2 to $m$ and are in excellent agreement with the theory \cite{Manan-Ko}. 

\section{Structure of Eigenstates: NPC, $l_H$ and entropy production $S(t)$}
\label{sec:4}

We utilize NPC (also known as participation ratio) and localization lengths $l_H$ defined by the information entropy ($S^{info}$) to analyze structure of eigenfunctions in many-body bosonic systems. 

Given the expansion of the many-particle basis states $\l.\l|k\r.\ran$, with energies $E_k = \langle k|H|k \rangle$, in the $H$ eigenvalue $E$ basis, $|k \rangle=\sum_E {C_{k}^{E}} \l.\l|E\r.\ran$, NPC, $S^{info}$ and $l_H$ are given as,
\be
\barr{rcl}
\mbox{NPC}(E) &=& \left\{ {\dis\sum\limits_k {\left| {C_{k}^{E} } \right|^4 } } \right\}^{ - 1} \;,\;\;\;
S^{info} (E) =  - \dis\sum_k {\left| {C_{k}^{E} } \right|^2 } \ln \left| {C_{k}^{E}} \right|^2\;,\\ \\
l_H(E)&=&\exp{S^{info} (E)/(0.48 \; d_m)} \;.
\earr
\label{eq.ijm-chv7}
\ee
NPC essentially gives the the number of basis states $\l.\l|k\r.\ran$ that constitute an eigenstate with energy $E$. Similarly, increase in information entropy implies more delocalization of the eigenstates. The GOE value for NPC is $d_m/3$ and for $\exp S^{info} = 0.48 \, d_m$.
	
\begin{figure}[thb]
		\begin{center}
\includegraphics[width=0.9\linewidth]{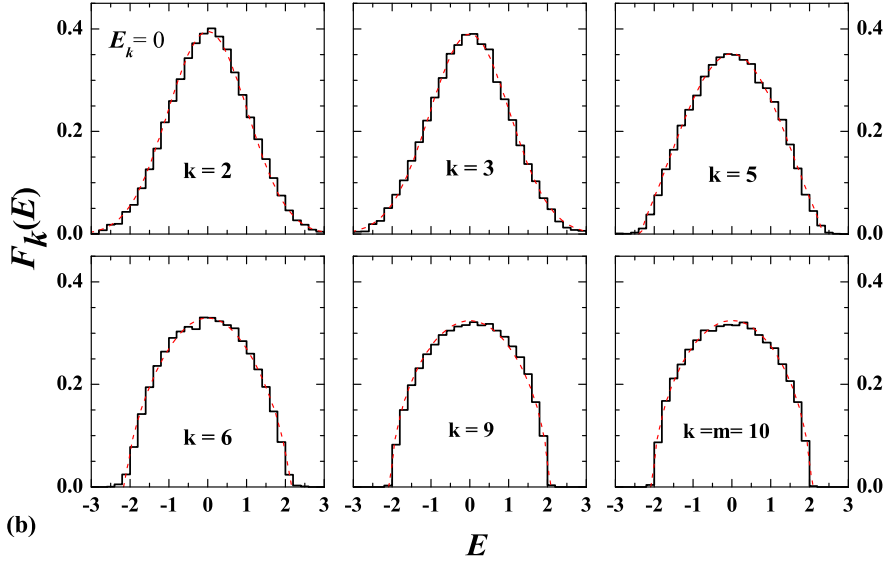}
		\end{center}
		\caption{Ensemble averaged LDOS as a function of normalized energies for a 100 member BEGOE(1+$k$) ensemble with $m = 10$ bosons in $N = 4$ sp states for different $k$ values. Here, interaction strength $\lambda = 0.5$ and $E_k=0$. The dashed (red) curves are obtained using Eq. \eqref{eq.ent7} with the corresponding values of $q$ given by Eq. \eqref{eq.ent9}.}
\label{fig.fke}
	\end{figure}

NPC and $S^{info}$ can be studied by examining the general features of LDOS (also known as strength functions) $F_k(E)$ that gives the spreading of the basis states over the eigenstates,
\be
F_{k}(E) = \sum_{E'} {|C_k^{E'}|}^2 \delta(E-E') = {|\cac_k^E|}^2 d_m \;\rho(E).
\ee
Here, ${|\mathcal{C}_k^E|}^2$ is the average of ${|C_k^E|}^2$ over the eigenstates with the same energy $E$ and $\rho(E)$ is the eigenvalue density. With the $E_k$ energies generated by a Hamiltonian $H_k$, it is easy to identify $F_k(E)$ as a conditional density of the bivariate density $\rho_{biv}(E,E_k) = \langle \delta (H-E)\rangle \langle \delta (H_k-E_k)\rangle$ given by \cite{kota-book},
\be
\barr{rcl}
\rho_{biv}(E,E_k) &=& \langle \delta (H-E)\rangle \langle \delta (H_k-E_k)\rangle=d_m^{-1} \dis\sum_{\alpha \in k, \beta \in E} {|C_{k,\alpha}^{E,\beta}|}^2 \\
&=& \dis\frac{1}{d_m} {|\mathcal{C}_k^E|}^2 [d_m \;\rho^H(E)] [ d_m \;\rho^H_k(E)]\;,\\\\
\rho_{biv}(E,E_k) &=& F_{k}(E) \rho^H_k(E)\;.
\earr
\ee
The above definition takes into account the degeneracy in the eigenspectrum. With $F_{k}(E)$ is a conditional density of the bivariate density $\rho_{biv}(E,E_k)$, the smooth forms for NPC$(E)$ and $l_H(E)$ can be written as,
\be
\barr{rcl}
\mbox{NPC}(E)&=&\dis\frac{d_m}{3} \rho(E)^2 \left\{ \dis\int dE_k \dis\frac{\rho_{biv}(E,E_k)^2}{\rho(E_k)} \right\}^{-1}\;,\\\\
l_H(E)&=& - \dis\int d E_k \dis\frac{\rho_{biv}(E,E_k)}{\rho(E)} \ln \left\{\dis\frac{\rho_{biv}(E,E_k)}{\rho(E)\rho(E_k)}\right\}\;.
\earr
\label{eq.ijm1}
\ee
Here, 
\be
\zeta=\sqrt{1-\dis\frac{\sigma_k^2}{\sigma_H^2}} 
\label{eq.zeta}
\ee
is the correlation coefficient of $\rho_{biv}(E,E_k)$; $\sigma_k^2$ is the variance of $\rho^H_k(E_k)$ and ${\sigma_H}^2$ is variance of $\rho^H(E)$. As the strength of the $k$-body interaction ($\lambda$) increases, value of $\zeta$ decreases.

For BEGOE(1+$k$) with sufficiently large interaction strength $\lambda$, ensemble averaged $F_k(E)$ changes from Gaussian to semi-circle form as the interaction rank $k$ changes from 2 to $m$ \cite{Manan-Ko}. Equation \eqref{eq.ent7} with the corresponding values of $q$ given by Eq. \eqref{eq.ent9} explains this transition very well \cite{Manan-Ko}. Figure \ref{fig.fke} shows an example for $F_k(E)$ for BEGOE(1+$k$) with $m=10$ bosons in $N=4$ sp states. There is a transition from Gaussian to semi-circle form in $F_k(E)$ as $k$ changes from 2 to $m$, just like spectral densities. The dashed (red) curves are obtained using Eq. \eqref{eq.ent7} with the corresponding values of $q$ given by Eq. \eqref{eq.ent9}. 

Taking $E_k$'s generated by $H_k= h(1)$, the form for $\rho_k^H(E_k)$ will be Gaussian. With this, it is possible to evaluate smooth forms for NPC$(E)$ and $l_H(E)$ for BEGOE(1+$k$) by replacing $q$-Hermite forms for the $F_k(E|q)$, $\rho_{biv:q}(E,E_k)$ and state density $\rho(E|q)$ in Eq. \eqref{eq.ijm1}. The $F_k(E|q)$ is the conditional density of bivariate $q$-normal $\rho_{biv:q}(E,E_k)$ given by \cite{SZAB},
\begin{equation}
\barr{l}
F_k(E|q) = \dis\frac{\sqrt{1-q}}{2\pi\sqrt{4-(1-q)E^2}} \times \\
\dis\prod_{k=0}^{\infty} \frac{(1-\zeta^2 q^k)(1-q^{k+1})((1+q^k)^2-(1-q)E^2q^k)}{(1-\zeta^2 q^{2k})^2-(1-q)\zeta q^k (1+\zeta^2 q^{2k}) E E_k + (1-q)\zeta^2(E^2+E_k^2)q^{2k}}\;,\\\\
 \rho_{biv:q}(E,E_k) = F_k(E|q) \;\rho_G(E_k)\;;\;\;\;\;
 \rho_G(E_k) = \dis\frac{1}{\sqrt{2\pi}} \exp (-E_k^2/2)\;.
\earr
\label{eq:qnormal}
\end{equation}
Here $\zeta$ is the correlation coefficient defined by Eq. \eqref{eq.zeta} and the centroids of the $E$ and $E_k$ energies are both given by $\epsilon_H=\langle H \rangle$. In principle, it is possible to obtain formulas for NPC and $l_H$ by substituting Eq. \eqref{eq:qnormal} in Eq. \eqref{eq.ijm1}. However, it is an open problem to simplify them. Therfore, we evaluate them numerically and compare it with the ensemble averaged results of BEGOE(1+$k$). 

\begin{figure}
 	\begin{center}
 			\includegraphics[width=\textwidth,height=3in]{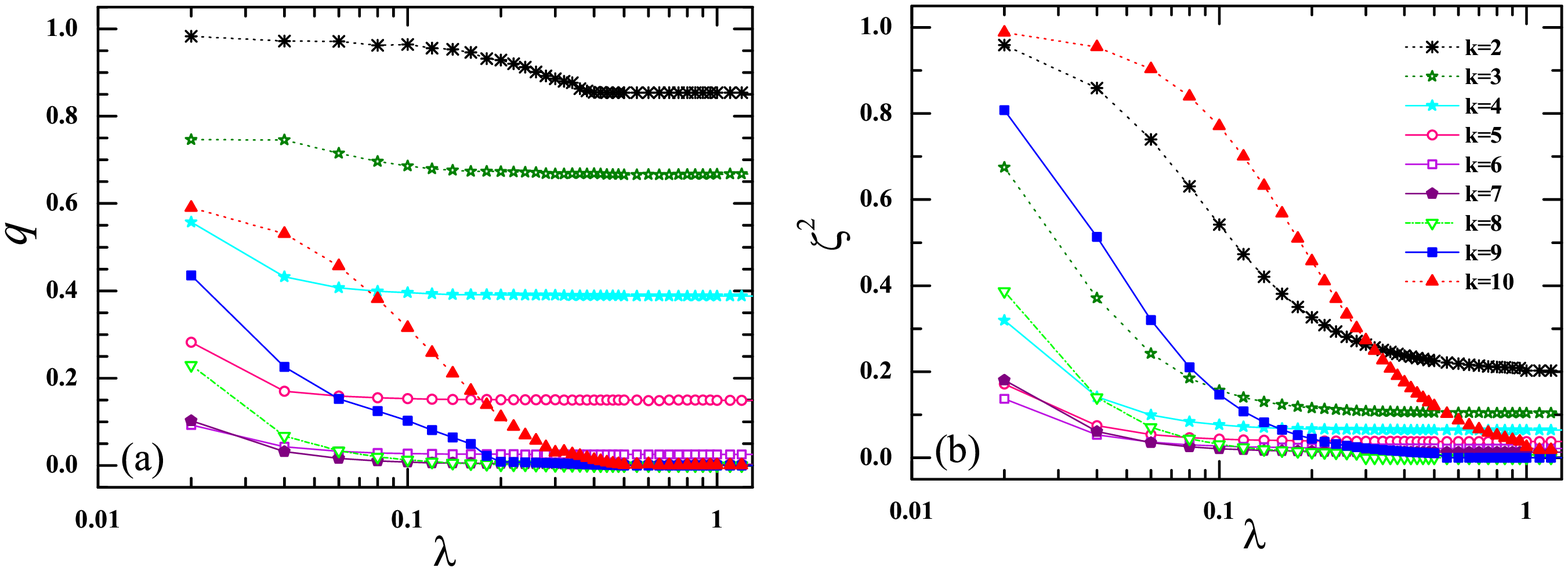}
 	\end{center}
 	\caption{For a 100 member BEGOE($1 + k$) ensemble with ($m = 10, N = 5$, $\lambda = 0.5$): Variation of (a) $q$ as a function of interaction strength $\lambda$ for different values of $k$; here, the $q$ values are extracted by fitting the ensemble averaged $\rho(E)$ given by Eq.~\eqref{eq.ent7}; and (b) ensemble averaged values of $\zeta^2$ as function of interaction strength $\lambda$ for different values of $k$.}
 	\label{fig-qzeta}	
 \end{figure}

For a 100 member BEGOE($1+k$) ensemble with $m=10$, $N=5$ and $\lambda = 0.5$, Figure \ref{fig-qzeta}(a) shows variation in $q$ values as a function of interaction strength $\lambda$ for different $k$ values. Here, the $q$ values are extracted by fitting the ensemble averaged $\rho(E)$ given by Eq.~\eqref{eq.ent7}. For a given $k$, $q$ value decreases as the strength of the interaction $\lambda$ increases and for sufficiently large $\lambda$, it approaches constant value close to given by Eq. \eqref{eq.ent9}. Hence, in the strong interaction limit, $k$-body part of the interaction dominates over one-body part and therefore BEGOE($1+k$) reduces to BEGOE($k$). Also, the value of $q$ for a given $\lambda$, decreases with increasing $k$ upto $k =6$ and then it increases again. Figure \ref{fig-qzeta}(b) shows the variation of $\zeta^2$ as a function of interaction strength $\lambda$ for a 100 member BEGOE(1+$k$) ensemble with 10 bosons in 5 sp states for different $k$ values. We choose $\lambda = 0.5$. For a given $k$, $\zeta^2$ decreases as $\lambda$ increases. Similar to $q$, the value of $\zeta^2$ decreases with increasing $k$ upto $k =6$ and then it increases again, for a given $\lambda$. 

 \begin{figure}[thb]
 	\begin{center}
 			\includegraphics[width=0.9\textwidth]{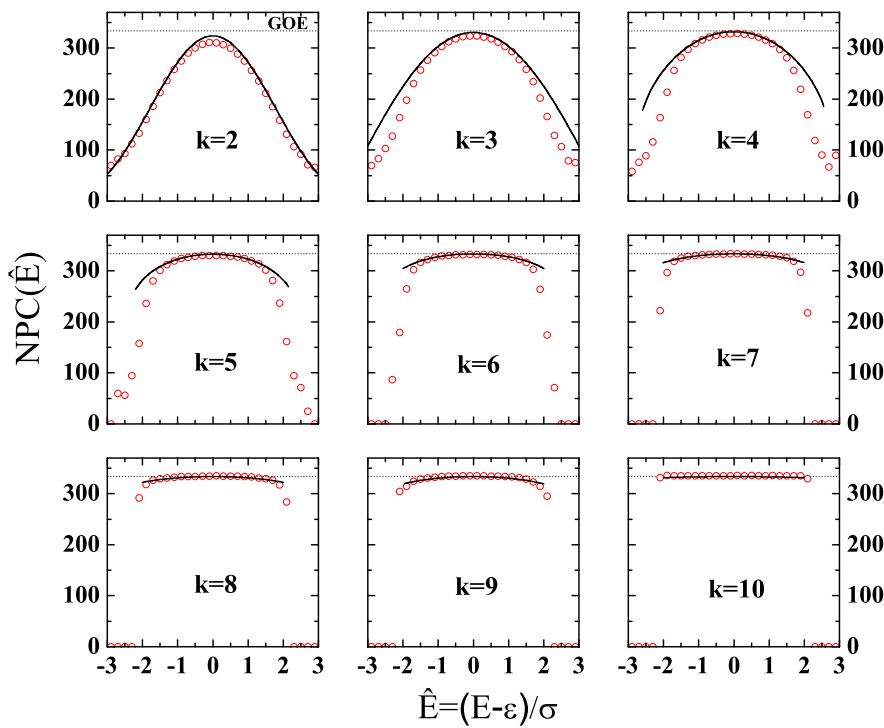}
 	\end{center}
 	\caption{Ensemble averaged NPC as a function of normalized energy $\hat{E}$ for a 100 member BEGOE(1+$k$) with 10 interacting bosons in 5 sp levels for different values of $k$. We choose $\lambda = $0.5 for $k=2-9$ and 1 for $k=m=10$. Numerical results (circles) are compared with the theoretical estimates (black curves) obtained using Eq. \eqref{eq.ijm1}.}
 	\label{fig-npc}	
 \end{figure}	

Fig. \ref{fig-npc}, shows the variation of ensemble averaged NPC for a 100 member BEGOE($1+k)$ with $m=10$ bosons in $N = 5$ sp states for different values of $k$. Here, $d_m = 1001$ and $\lambda$ is chosen to be 0.5 for $k=2-9$ and 1 for $k=m=10$. Numercal results (circles) are compared with theoretical estimates (black curves) obtained using Eqs. {\eqref{eq.ijm1} and \eqref{eq:qnormal}. The $\zeta$ values are computed from BEGOE(1+$k$) Hamiltonians using Eq. \eqref{eq.zeta} and the $q$ values are obtained using Eq. \eqref{eq.ent9}. We find excellent agreement between numerics and theory in the bulk of the spectrum.

\begin{figure}[thb]
 	\begin{center}
 			\includegraphics[width=0.9\textwidth]{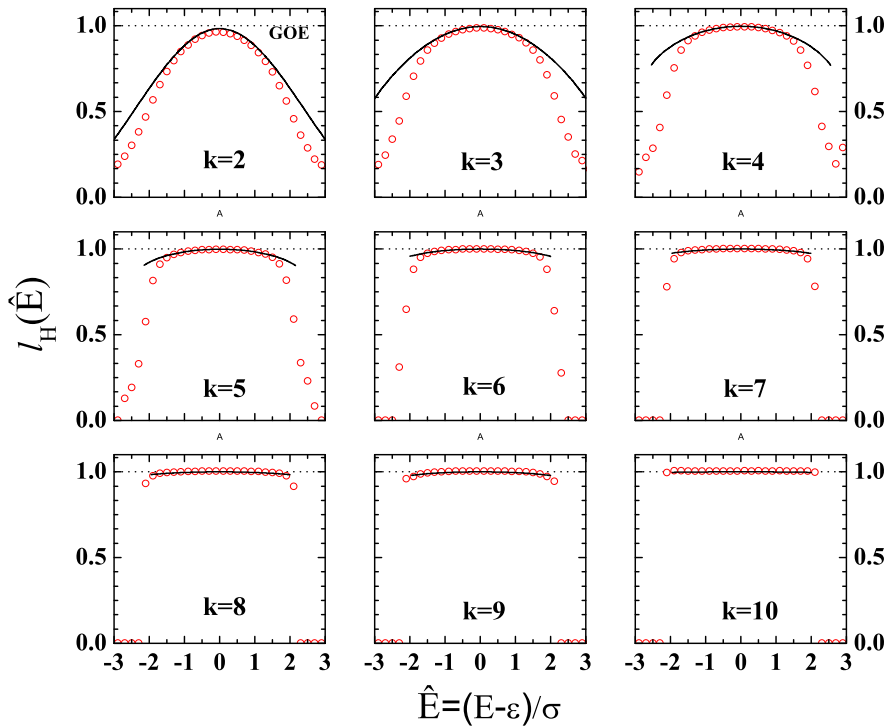}
 	\end{center}
 	\caption{Ensemble averaged localization lengths $l_H$ as a function of normalized energy $\hat{E}$ for a 100 member BEGOE(1+$k$) with 10 interacting bosons in 5 sp states for different values of $k$. Numerical results (circles) are compared with the theoretical estimates (black curves) obtained using Eq. \eqref{eq.ijm1}.}
 	\label{fig-lh} 	
 \end{figure}	

Similarly, Fig. \ref{fig-lh} compares the numerical results of $l_H$ with the expression given by Eq. \eqref{eq.ijm1} for different $k$ values. We obtain excellent agreement between numerical results (circles) and theory (black curves) for all values of $k$ in the bulk of the spectrum.

Next, we study entropy production $S(t)$ with time whose behavior is dictated by survival probability $F(t) = | \int \mbox{LDOS} \exp(-iEt) dE |^2$. The Fourier transform of generating function of $q$-Hermite polynomials explains the survival probability decay \cite{Manan-Ko}. In terms of $F(t)$, for small times, $S(t)$ can be expanded as \cite{Haldar},
\be
\barr{l}
S(t) \approx -F(t) \ln F(t) - \l[1-F(t)\r] \ln \l(\dis\frac{1-F(t)}{n}\r)\;;\\
n \sim \alpha (\rm{NPC}_{max}) = \alpha \dis\frac{d_m}{3} \dis\sqrt{1-\zeta^4}\;.
\earr \label{eq.time5}
\ee
Here, $\alpha$ is a free parameter. Figure \ref{ent-prod} shows the entropy $S(t)$ for a 100 member BEGOE($1+k$) ensmeble with ($m = 10$, $N = 4$, $\lambda = 0.5$) for selected $k$ values. Ensemble averaged $S(t)$ values (circles) are compared with curves (red) obtained using Eq. (\ref{eq.time5}). Agreement between numerics and theory gets better with increasing $k$. 

	\begin{figure}[t]
		\begin{center}
			\includegraphics[width=0.9\linewidth]{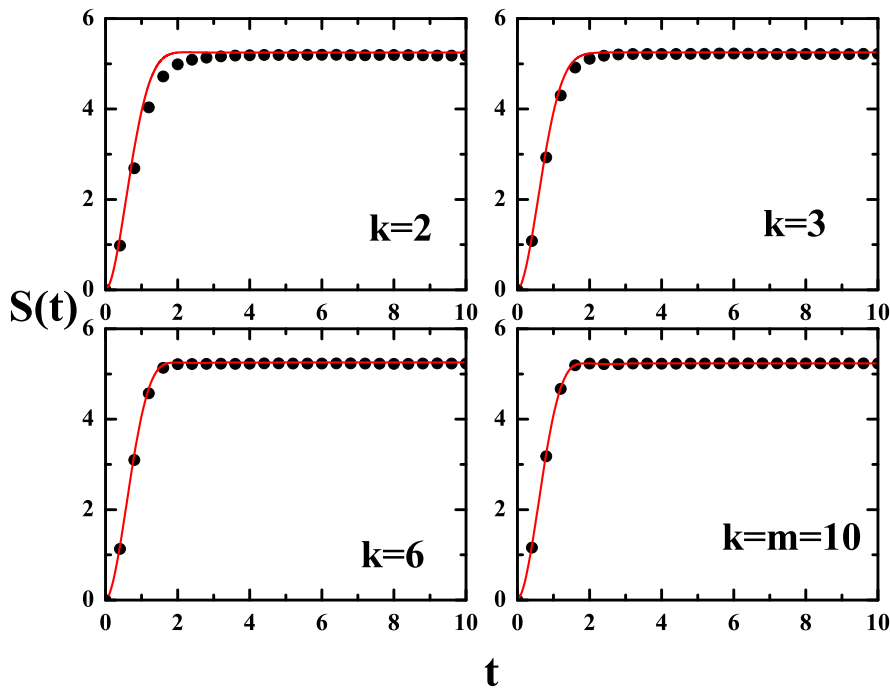}
		\end{center}
		\caption{Entropy $S(t)$ as a function of time for various values of $\lambda$ for a 100 member BEGOE(1+$k$) ensemble with $N = 4$, $m = 10$. The open circles are ensemble averaged results while smooth curves are obtained using Eq. (\ref{eq.time5}).}
\label{ent-prod}
	\end{figure} 

Realistic system preserve additional symmetries (in addition to particle number $m$) like angular momentum, parity, spin-isospin $SU(4)$ symmetry, and so on \cite{kota-book,Manan-thesis}. Transport efficiencies get enhanced in quantum systems modeled by embedded ensembles with centrosymmetry \cite{centro-1,centro-2}. In the next section, we compare results for transport efficiencies using BEGOE($k$) model with and without centrosymmetry.

\section{Transport efficiencies using BEGOE($k$): Role of centrosymmetry}
\label{sec:5}

\begin{figure}
 	\begin{center}
 			\includegraphics[width=0.9\textwidth]{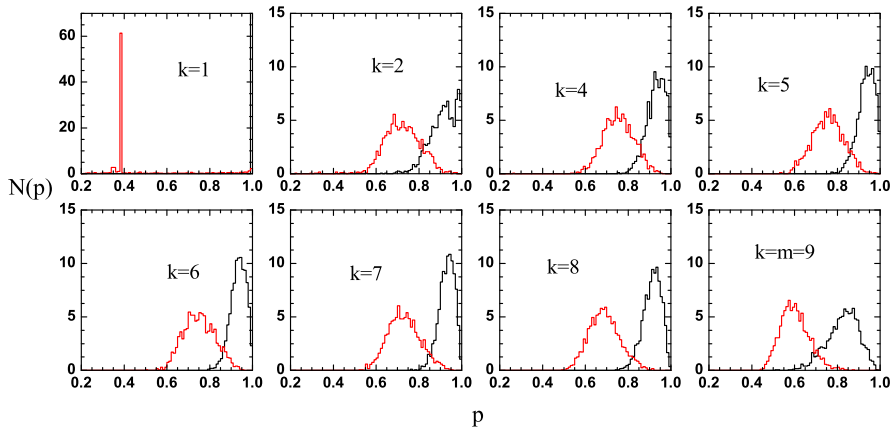}
 	\end{center}
 \caption{Normalized distributions of the best efficiencies $P$ for a 2000 member (i) BEGOE($k$) (red histogram), (ii) BEGOE($k$-cs) (green histogram) and (iii) csBEGOE($k$) (black histogram)  with $N = 2$ and $m=9$ as a function of interaction rank $k$. Cases (ii) and (iii) are identical for $N = 2$.}
 	\label{fig-92}
\end{figure}

Efficient transport of particles or excitations (known as quantum efficiency) within a complex quantum system is a very important as well as a challenging task. Its understanding is essential in a variety of fields like quantum information science, photosynthesis, solar cell physics etc \cite{centro-appl}. 

In realistic complex quantum systems, only a limited degree of control is available. Hence, it is important to find out under what conditions near-to-perfect transport between two states of a small disordered
interacting quantum system can be improved. Using a disordered network of $d_m$ sites and employing GOE for the Hamiltonian, it was shown in \cite{centro-prl} that highly efficient quantum transport is possible with centrosymmetry of $H$ and a dominant doublet spectral structure. Following this study, it was shown that the transport efficiency is enhanced just with centrosymmetry when GOE is replaced by embedded GOE  \cite{centro-1}.

Modeling a complex network Hamiltonian $H$ with BEGOE($k$) of $m$ bosons distributed in $N$ sp states, the basis states generating the Hilbert space (of dimension $d_m$) are treated as nodes of the network. Initially, the network is prepared in state $\ket{in} = \ket{i}$ and an excitation is introduced. As the network evolves unitarily, the excitation propagates to the state $\ket{out} = \ket{f}$. Note that the the initial
$\l.\l|i\r.\ran$ and final $\l.\l|f\r.\ran$ excitations are localized on the nodes of the network. Then, the maximum transition probability achieved among theses states within a time interval $[0, T]$ is termed as the transport efficiency, which is quantitatively defined as \cite{centro-prl,Zech2014}
\be
P_{i,f} = max_{[0,T]}  \l|\lan f |U(t)| i \ran\r|^2\;.
\label{eq.cent-1}
\ee
Here, $U(t)$ is the unitary quantum evolution associated with the Hamiltonian $H$ of the network. The network is said to have perfect state transfer (PST) when $P_{i,f} = 1$. 

Centrosymmetry is defined by the condition $J H = H J$ where $J$ is the exchange  matrix with $J_{i,j}=\delta_{i,d-i+1}$. Using Embedded Gaussian Orthogonal and Unitary Ensembles for $H$ with centrosymmetry, transport efficiencies given by Eq. (\ref{eq.cent-1} have been investigated for both fermionic and bosonic systems \cite{centro-1}. 

We compare transport efficiencies by employing three models: (i) BEGOE($k$), (ii) BEGOE($k$) with centrosymmetry present in $k$ particle spaces (not in the $m$-particle spaces) [denoted as BEGOE($k$-cs], and iii) BEGOE($k$) with centrosymmetry present both in $k$ and $m$ particle spaces [denoted as csBEGOE($k$)]. BEGOE($k$-cs) ensemble is constructed by imposing centrosymmetry in $k$-particle spaces and then propagating it to $m$ particle spaces using the many-particle Hilbert space geometry. The final Hamiltonian will not preserve the centrosymmetry structure in this case. While csBEGOE($k$) is constructed by imposing centrosymmetry in the one particle space and propagating it to $k$ and $m$ particle spaces \cite{centro-1,centro-2}. 
  	
 \begin{figure}
 	\begin{center}
 			\includegraphics[width=\textwidth,height=5in]{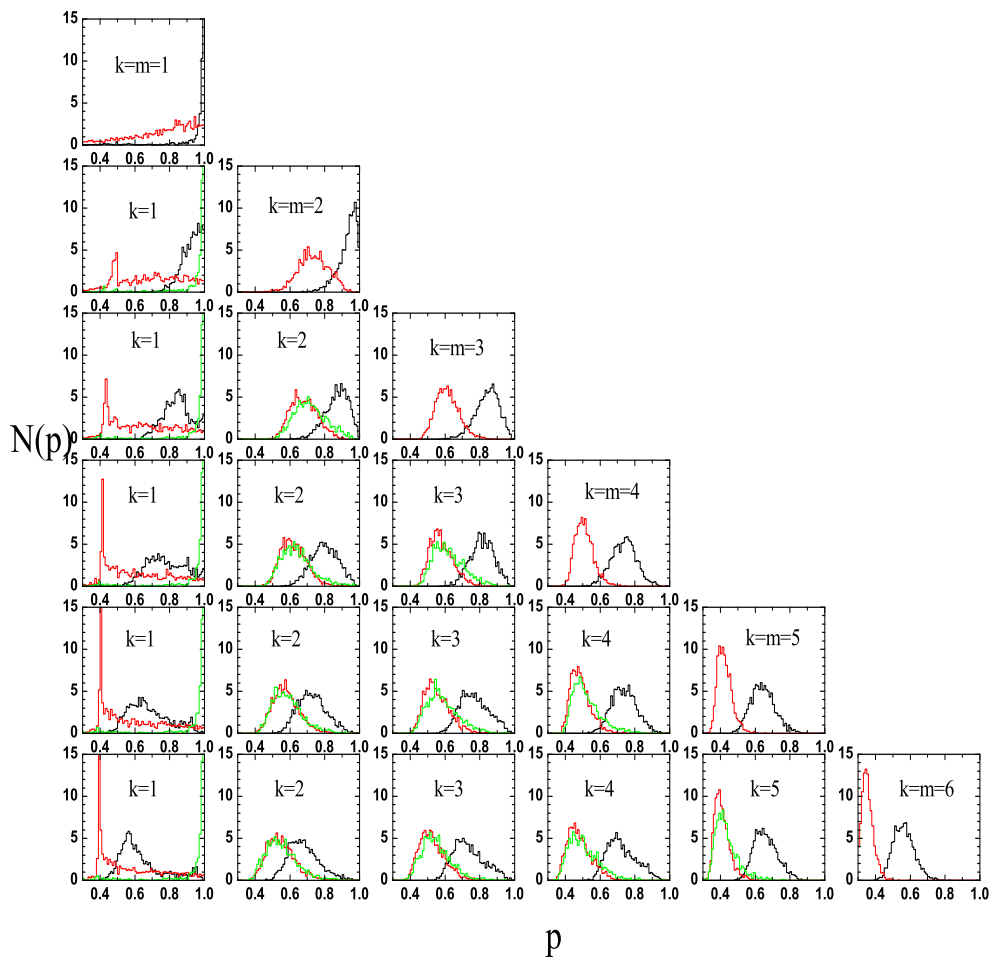}
 	\end{center}
 \caption{Normalized distributions of the best efficiencies $P$ for a 2000 member (i) BEGOE($k$) (red histogram), (ii) BEGOE($k$-cs) (green histogram) and (iii) csBEGOE($k$) (black histogram) with $N = 3$ and $m=6$ as a function of $m$ and interaction rank $k$. Rows have the same particle number $m$ and columns the same $k$ value. Note that for $k=m$, results for BEGOE($k$-cs) and csBEGOE($k$) are identical. }
 	\label{fig-N3}
  \end{figure}	

We consider a network generated by basis states obtained by distributing $m = 9$ bosons in $N = 2$ sp levels. The total number of basis states are $10$ in this case and we represent the network Hamiltonian by a 2000 member (i) BEGOE($k$), (ii) BEGOE($k$-cs) and (iii) csBEGOE($k$). Distributions of the best efficiencies $P$ of each member of ensemble is calculated as a function of interaction rank $k$ and the normalized distributions are shown in Fig. \ref{fig-92}. The red histograms correspond to BEGOE($k$) and the black histograms correspond to csBEGOE($k$). For a two-level ($N = 2$) BEGOE($k$), csBEGOE($k$) and BEGOE($k$-cs) are identical by construction \cite{centro-1}. It is evident from these results that centrosymmetry enhances transport efficiencies.

Next, we consider a network generated by basis states obtained by distributing $m = 6$ bosons in $N = 3$ sp levels. The total number of basis states are $28$ in this case and we represent the network Hamiltonian by a 2000 member (i) BEGOE($k$) (red histogram), (ii) BEGOE($k$-cs) (green histogram) and (iii) csBEGOE($k$) (black histogram). Normalized distributions of the best efficiencies $P$ are shown in Fig. \ref{fig-N3}. In the figure, the number of bosons $m$ is same along the rows while the value of $k$ is same along the columns. It can be seen that the transport efficiency for BEGOE($k$) is less than 80\% for almost all the cases while with BEGOE($k$-cs) there is a marginal improvement for all $k>1$. On the other hand, with csBEGOE($k$), one can observe that the efficiency is around 80\%, which demonstrates that the presence of centrosymmetry enhances the transport efficiency. Note that for $k=m$, BEGOE($k$-cs) and csBEGOE($k$) are identical. For csBEGOE($k$), there is a PST for $m=3$ and $k \leq 3$. It is interesting to note that for $k = m = 1$, BEGOE($k$-cs) gives PST for $m = 1-6$. The lack of PST for $N = 3$ levels beyond $m = 3$ and $k \leq 3$ in comparison to $N=2$ example can be attributed to a systematic appearance of doublets in the spectrum for $N = 2$ \cite{Christandl2005}.

\section{Conclusions}
\label{sec:6}

We have analyzed structure of eigenfunctions in many-body bosonic systems by modeling the Hamiltonian of these complex systems using BEGOE(1 + $k$). The quantities employed are the number of principal components (NPC), the localization length ($l_H$) and the entropy production $S(t)$. The numerical results are compared with the analytical formulas obtained using random matrices which are based on bivariate $q$-Hermite polynomials for LDOS $F_k(E|q)$ and the bivariate $q$-Hermite polynomial form for bivariate eigenvalue density $\rho_{biv:q}(E,E_k)$ that are valid in the strong interaction domain.  We have studied variation in $q$ value as a function of interaction strength $\lambda$ for a fixed $k$ and found that in the strong interaction limit, $k$-body part of the interaction dominates over one-body part and therefore,  BEGOE(1+$k$) reduces to BEGOE($k$). The bivariate correlation coefficient $\zeta^2$ (which measures the variance of the distribution of centroids of LDOS relative to the variance of spectral density) decreases as the interaction strength $\lambda$ increases. We also compare transport efficiency in many-body bosonic systems using BEGOE in absence and presence of centrosymmetry. It is seen that the centrosymmetry enhances quantum efficiency. In future, it will be interesting to examine the transport efficiencies by imposing different centrosymmetries on embedded ensembles.

\section{Acknowledgments}
\label{sec:7}

It is a pleasure to thank V. K. B. Kota for useful discussions. MV acknowledges financial support from UNAM/DGAPA/PAPIIT research Grant IA101719.

\end{document}